\documentclass[a4paper,11pt]{article}
\usepackage{jcappub} 
\usepackage{orcidlink}
\usepackage{xcolor}  
\usepackage{makecell}
\usepackage{amsmath}
\usepackage{longtable}
\usepackage{multirow}
\usepackage{romannum}
\usepackage{booktabs}  
\usepackage{threeparttable}  
\usepackage{adjustbox}

\title{Forecasting Constraint on Primordial Black Hole Properties with the CSST \(3\times2\)pt Analysis}

\author[a, b]{Dingao Hu~\orcidlink{0009-0009-2259-5221},}
\author[a,b,c,1]{Yan Gong~\orcidlink{0000-0003-0709-0101},\note{Corresponding author.}}
\author[a, b]{Pengfei Su~\orcidlink{0009-0007-5468-8241},}
\author[a]{Hengjie Lin~\orcidlink{0000-0003-4528-656X},}
\author[a]{Haitao Miao~\orcidlink{0000-0003-0850-3641},}
\author[a, b]{Qi Xiong~\orcidlink{0009-0009-1369-2476},}
\author[a, b, d, e, f]{and Xuelei Chen~\orcidlink{0000-0001-6475-8863}}

\affiliation[a]{National Astronomical Observatories, Chinese Academy of Sciences, Beijing, 100101, China}
\affiliation[b]{School of Astronomy and Space Sciences, University of Chinese Academy of Sciences, Beijing, 100049, China}
\affiliation[c]{Science Center for Chinese Space Station Survey Telescope, National Astronomical Observatories, Chinese Academy of Science, Beijing 100101, China}
\affiliation[d]{Department of Physics, College of Sciences, Northeastern University, Shenyang 110819, China}
\affiliation[e]{Centre for High Energy Physics, Peking University, Beijing 100871, China}
\affiliation[f]{State Key Laboratory of Radio Astronomy and Technology, China}

\emailAdd{huda@bao.an.cn} 
\emailAdd{gongyan@bao.ac.cn}
\emailAdd{supf@bao.ac.cn}
\emailAdd{xuelei@bao.ac.cn}

\abstract{This study forecasts the constraints on the properties of primordial black holes (PBHs) as a cold dark matter component using the galaxy clustering, weak lensing, and galaxy-galaxy lensing (i.e. \(3\times2\)pt) measurements from the upcoming Chinese Space Station Survey Telescope (CSST) photometric survey. Since PBHs formed via gravitational collapse in the early Universe, they can additionally affect the formation and evolution of the cosmic large-scale structure (LSS) through  ``Poisson" effect. We compute the angular power spectra for PBH-\texorpdfstring{$\Lambda$}{Lambda}CDM cosmology, and generate mock data based on the CSST instrumental and survey design. The Markov Chain Monte Carlo (MCMC) method is employed to constrain the free parameters, such as the product of the PBH fraction and mass $f_{\rm PBH}m_{\rm PBH}$ and other cosmological parameters. The systematic parameters are also included in the fitting process, such as the parameters of baryonic effect, intrinsic alignment, galaxy bias, photometric redshift (photo-$z$) calibration, shear calibration, and noise terms. We find that the CSST 3$\times$2pt analysis can achieve tight constraints on $f_{\rm PBH}m_{\rm PBH}$, with 68\% and 95\% confidence levels (CLs) reaching $<10^{3.9} M_{\odot}$ and $<10^{4.7} M_{\odot}$, respectively. Additionally, the cosmological parameters, e.g. $\Omega_m$, $\sigma_8$ and $w$, can be constrained with the precisions of \(3.3\%\), \(1.7\%\), \(13\%\), respectively. This indicates that the CSST 3$\times$2pt analysis is a powerful tool to advance the PBH dark matter studies in the near future. }

\begin{document}
\maketitle
\flushbottom

\section{Introduction}\label{sec: Intro}
Primordial black holes (PBHs) are believed to form through gravitational collapse of large overdensity regions in the early Universe. Various  theories describe the PBH formation, including inhomogeneities during the radiation-dominated era \citep{carr_primordial_1975}, critical collapse \citep{kuhnel_effects_2016,carr_primordial_2021}, etc. Depending on the PBH formation time from \(10^{-43}\) to 1 s, the PBH mass ranges from \(10^{-5}\) gram to $10^{5}\,M_{\odot}$. In some theories, the mass of PBHs can even extend to $10^{22}\,M_{\odot}$ \citep{Carr:2024nlv}. Due to their early formation and wide mass range, PBHs can explain various observational effects as cold dark matter (CDM), such as the cross-correlations between the fluctuations in the source-subtracted Cosmic Infrared Background (CIB) and the unresolved Cosmic X-ray Background (CXB) \citep{Kashlinsky:2016sdv,Cappelluti:2012mj}, the UV luminosity density of James Webb Space Telescope (JWST) \citep{matteri_beyond_2025}, etc. Different methods are used to constrain the properties of PBHs for different mass ranges, including \(\gamma\)-ray background \citep{Arbey:2019vqx} and anisotropies of cosmic microwave background (CMB) \citep{serpico_cmb_2024} for small-mass PBHs, microlensing for solar-mass PBHs  \citep{carr_constraints_2021}, and dynamical motions, Lyman-\(\alpha\) forest \citep{Ivanov:2025pbu}, cosmic large-scale structure (LSS) \citep{Byrnes:2025tji}, and the combination of CMB and Baryonic Acoustic Oscillation (BAO) \citep{Gerlach:2025vco} for large-mass PBHs.

The combination of galaxy clustering and cosmic shear measurements, i.e. $3\times2$pt, is a powerful probe for the LSS, which is widely used to constrain the cosmological parameters \citep{hu_joint_2004, DES:2021wwk}, including the Kilo-Degree Survey (KiDS)\citep{Stolzner:2025htz, Wright:2025xka}, Dark Energy Survey (DES) \citep{DES:2017myr, DES:2021wwk}, Hyper Suprime-Cam (HSC) \citep{PhysRevD.108.123520, PhysRevD.108.123521}, etc. Especially, the upcoming Stage~\Romannum{4} surveys can cover larger redshift range and sky area, and will constrain cosmological parameters more tightly, include the Chinese Space Station Survey Telescope (CSST) \citep{zhan_consideration_2011, zhan_wide-field_2021, gong_cosmology_2019,gong_future_2025,collaboration_introduction_2025}, Vera Rubin Observatory's Legacy Survey of Space and Time (LSST) \citep{Ivezic_2019}, $\it Euclid$ \citep{Euclid:2025hny, Euclid:2025pzh}, and Nancy Grace Roman Space Telescope (RST)\citep{green_wide-field_2012}.

Currently, due to the limited sky area and depth of the Stage~\Romannum{3} surveys, there is no effective constraint on the primordial black hole parameters using the $3\times2$pt probe. In this work, we forecast the CSST $3\times2$pt probe to constrain the PBH properties. The CSST is a 2-meter space telescope and will co-orbits with the China Manned Space Station. It is planned to launch around 2027, and will cover a sky area of 17500 deg$^2$ in about ten years with field of view of 1.1 deg$^2$. It is expected to obtain more
than one billion galaxy images and one hundred million galaxy spectra in a redshift range $z<4$. The CSST can effectively constrain the dark energy and dark matter properties \citep{gong_cosmology_2019,xiong_cosmological_2024,miao_forecasting_2024,song_cosmological_2024,Song:2024esa}, neutrino mass \citep{Lin:2022aro}, ultralight axions \citep{Lin:2023yso}, modified gravity models\citep{Chen:2022wll,Yan:2024jwz} , etc. The great observational ability of the CSST will also has significant potential for constraining the PBH properties.

The paper is organized as follows : in Section~\ref{sec: angular power spectra}, we discuss the PBH effects in cosmology and the theory of angular power spectra for galaxy clusters and cosmic shear; in Section~\ref{sec: mock data}, we introduce the method of generating and selecting the mock data of angular power spectra; in Section \ref{sec: result}, we discuss the \(3\times2\)pt fitting method and constraint results of the free parameters; Section \ref{sec: summary} gives the summary and relevant discussions. We assume a flat $\Lambda$CDM cosmology with the fiducial values of the cosmological parameters from {\it Planck} measurements \citep{Planck:2018vyg}.

\section{Theoretical Model}\label{sec: angular power spectra}
\subsection{PBH-$\Lambda$CDM power spectrum}
PBHs can be seen as a component of CDM, and could cause specific effects on structure formation through their isocurvature perturbations. Therefore, the properties of PBHs can be constrained by the measurements of cosmic structure formation. The influence of PBHs can be specifically categorized into two types: ``Poisson" effect and ``Seed" effect. The ``Poisson" effect represents the collective influence of multiple PBHs coexisting within a given region, and the ``Seed" effect represents the isolated influence of a single PBH within a given region \citep{carr_primordial_2018}. Both of the two effects can basically enhance the matter power spectrum at small scales. However, since there is still no reliable model that can accurately describe the ``Seed" effects, current studies typically apply a cut-off at small scales in the matter power spectrum to obtain conservative results  \citep{liu_accelerating_2022,Gouttenoire:2023nzr,Hutsi:2022fzw}. In this work, we also adopt the cut-off and only consider the power spectra in the linear or quasi-linear regime as described in the next section, where the ``Seed" effect can be safely neglected. Regarding the mass spectrum of PBHs, it has shown that constraints on a monochromatic mass spectrum can be extended to arbitrary mass spectra \citep{carr_primordial_2017}. Therefore, in this work, we assume that PBHs have a monochromatic mass spectrum. 

\begin{figure}
    \centering
    \includegraphics[width=0.6\linewidth]{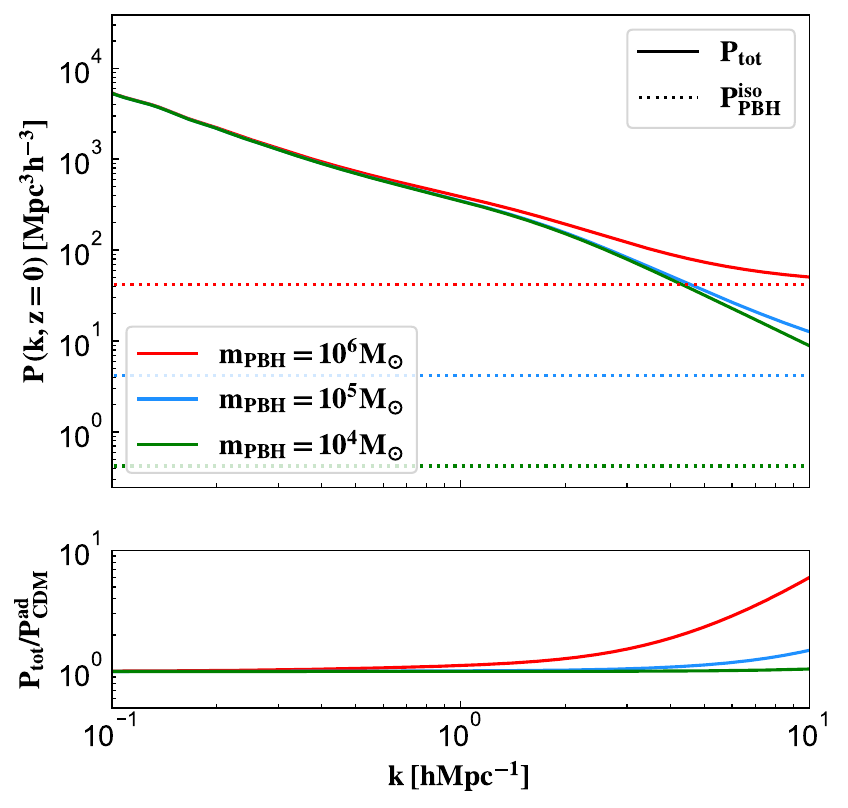}
    \caption{The total PBH-\(\Lambda\)CDM power spectra (solid curves) and PBH isocurvature power spectra (dotted lines) in the PBH-\(\Lambda\)CDM model at \(z = 0\) with $f_{\rm PBH} = 1$ for the  three PBH masses, i.e. $m_{\rm PBH}=10^4$ (green), $10^5$ (blue), and $10^6M_{\odot}$ (red). The ratio of the PBH-\(\Lambda\)CDM and \(\Lambda\)CDM matter power spectrum for each $m_{\rm PBH}$ is also shown in the lower panel. 
    \label{fig: matter power spectrum}}
\end{figure}

Under these conditions, the initial matter power spectrum for PBH isocurvature perturbations is given by \citep{afshordi_primordial_2003} 
\begin{equation}
    P^{\rm iso}_{\rm PBH}=f_{\rm PBH}^2\,\overline{n}_{\rm PBH}^{-1}=\frac{8\pi G\,f_{\rm PBH}\,m_{\rm PBH}}{3H_0^2\,\Omega_{\rm CDM}},
    \label{equ:PBH power spectrum }
\end{equation}
where $f_{\rm PBH}$ is the PBH fraction of CDM, $m_{\rm PBH}$ is the mass of a single PBH, and \(\overline{n}_{\rm PBH}\) is the mean volume number density of PBHs in comving coordinate system. Since $f_{\rm PBH}$ and $m_{\rm PBH}$ are completely degenerate, the product $f_{\rm PBH}\,m_{\rm PBH}$ is treated as a free parameter in our analysis. $\Omega_{\rm CDM}=\Omega_m-\Omega_b$ is the CDM density parameter, where $\Omega_m$ and $\Omega_b$ are the matter and baryon density parameters, respectively. The gravitational constant is \(G=6.67\times10^{-11}\,\rm m^3kg^{-1}s^{-2}\) and the Hubble parameter is \(H_0\equiv100\,h\,\rm km\,s^{-1}Mpc^{-1}\) with \(h=0.6727\). Note that the power spectrum of the PBH adiabatic perturbation $P^{\rm ad}_{\rm PBH}$ is identical to that of ordinary CDM, which can be absorbed into the matter power spectrum $P^{\rm ad}_{\rm CDM}$ of the standard $\Lambda$CDM \citep{afshordi_primordial_2003,inman_early_2019}.

Considering the evolution of PBH isocurvature perturbations and the \(\Lambda\)CDM standard adiabatic model, the total power spectrum of the PBH-\(\Lambda\)CDM cosmology can be expressed as
\begin{equation}
    P_{\rm tot}(k,z)= P^{\rm ad}_{\rm CDM}(k,z)+D^{\rm iso}(z)^2P^{\rm iso}_{\rm PBH},
    \label{equ:tatol matter power spectrum}
\end{equation}
where \(D^{\rm iso}(z)\) is the growth factor for PBH isocurvature perturbations, follows the expression as below \citep{inman_early_2019}
\begin{equation}
  \begin{aligned}
      D^{\rm iso}(z) & \approx (1+\frac{3\gamma}{2\alpha_-}s)^{\alpha_-},\quad s= \frac{1+z_{\rm eq}}{1+z},\\
      \alpha_- & =\frac{1}{4}(\sqrt{1+24\gamma}-1),\gamma=\frac{\Omega_{\rm CDM}}{\Omega_m},
      \label{equ:growth factor}
  \end{aligned}
\end{equation}
where \(z_{\rm eq} = 3400\) is the redshift at matter-radiation equality. Since PBHs mainly affect the cosmic structures at small scales, we estimate the non-linear matter power spectrum $P^{\rm ad}_{\rm CDM}(k,z)$, which incorporates the baryonic effect parameter \({\rm log}_{10}(T_{\rm AGN}/K)\), by using \href{https://camb.readthedocs.io/en/latest/index.html}{\tt CAMB} \citep{Lewis:1999bs} with HMCode-2020 \citep{Mead:2020vgs}. We then apply this  power spectrum to the subsequent calculations for all three cosmological probes.

In Figure \ref{fig: matter power spectrum}, we show the total (solid) and PBH (dotted) isocurvature perturbations power spectra in the PBH-\(\Lambda\)CDM cosmology for $m_{\rm PBH}=10^4$, $10^5$, and $10^6$ $\rm M_{\odot}$ with \(f_{\mathrm{PBH}}=1\) at $z=0$. We can find that the PBH power spectrum is a constant for each  $m_{\rm PBH}$, and can significantly affect the total power spectrum at small scales of the non-linear regime with $k\gtrsim0.7$ and $4\ {\rm Mpc^{-1}}h$ for $m_{\rm PBH} = 10^6$ and $10^5\,M_{\odot}$, respectively. As a simple order-of-magnitude estimate, by varying the value of parameter \(f_{\rm PBH}\,m_{\rm PBH}\) and following the \(\chi^{2}\) estimation described in Section 4, we can find that the value of \(f_{\rm PBH}\,m_{\rm PBH}\) is \(\sim10^{4} M_{\odot}\) when \(\Delta\chi^{2} \sim O(1)\).

\subsection{Galaxy angular power spectrum}

After obtaining the total matter power spectrum in the PBH-\(\Lambda\)CDM cosmology, we can estimate the galaxy angular auto- or cross-power spectrum for the $i$-th and $j$-th tomographic redshift bins, which is given by
\begin{equation}
    \tilde{C}_{\mathrm{gg}}^{ij} = C_{\mathrm{gg}}^{ij} + \frac{\delta_{ij}}{\bar{n}_{\rm g}^{i}} + N_{\mathrm{sys,g}}^{\mathrm{ij}}.
    \label{galaxy angular power spectrum observed by CSST }
\end{equation}
Here \(\delta_{ij}/\bar{n}_{\rm g}^{i}\) represents the shot noise term, where \(\delta_{ij}\) is Kronecker delta function and $\bar{n}_{\rm g}^{i}$ is the mean galaxy surface number density in the $i$-th bin. \(N_{\rm sys,g}^{\rm ij}\) denotes the systematic error term for galaxy clustering survey, which should be redshift-dependent. For simplicity, we adopt a constant value of \(N_{\rm sys,g}^{\rm ij}=N_{\rm sys}^{\rm g}\) for all redshift bins, and we take \(N_{\rm sys}^{\rm g}=10^{-8}\) as the fiducial value \citep{gong_cosmology_2019}. The quantity \(C_{\rm gg}^{ij}\) is the clustering term of the galaxy angular power spectrum. Using the flat sky assumption and Limber approximation \citep{Limber:1954zz}, it can be computed as \citep{hu_joint_2004}
\begin{equation}
    C_{\mathrm{gg}}^{ij}(\ell) =\int dz \,H(z) \frac{W_{\mathrm{g}}^{i}(z) W_{\mathrm{g}}^{j}(z)}{cD_{\mathrm{A}}^{2}(z)} P_m\left( \frac{\ell + \frac{1}{2}}{D_{\mathrm{A}}(z)}, z \right),
\end{equation}
where $c$ is the speed of light, \(H(z)\) is the Hubble parameter, \(D_{\rm A}\) is the comoving angular diameter distance, $P_m(k,z)=P_{\rm tot}(k,z)$ is the total matter power spectrum in the PBH-\(\Lambda\)CDM cosmology in our analysis. \(W_{\mathrm{g}}^{i}(z)\) is the galaxy weight kernel for the $i$-th bin, which is expressed as
\begin{equation}
    W_{\mathrm{g}}^{i}(z) = b^{i}_{\rm g}(z)\, n_{\rm g}^{i}(z). \label{equ: galaxy weight function}
\end{equation}
Here $b^{i}_{\rm g}(z)$ is the galaxy bias. Note that the galaxy bias should be scale-dependent and may affect the clustering at small scales. Since we only focus on the linear regime in the CSST galaxy clustering survey, for simplicity, we assume that $b^{i}_{\rm g}$ is a scale-independent constant within each redshit bin \citep{gong_cosmology_2019}. The fiducial value is calculated by $b^{i}_{\rm g}(z)=1+0.84z_{\rm cen}^i$, where $z_{\rm cen}^i$ is the central redshift of the $i$-th bin \citep{weinberg_galaxy_2004}. In the fitting process, we set it as a free parameter in each tomographic bin. $n_{\rm g}^{i}(z)$ is the normalized galaxy redshift distribution in the $i$-th bin, which has $\int n_{\rm g}^{i}(z) dz=1$.
\begin{figure*}
    \centering
    \includegraphics[width=0.49\linewidth]{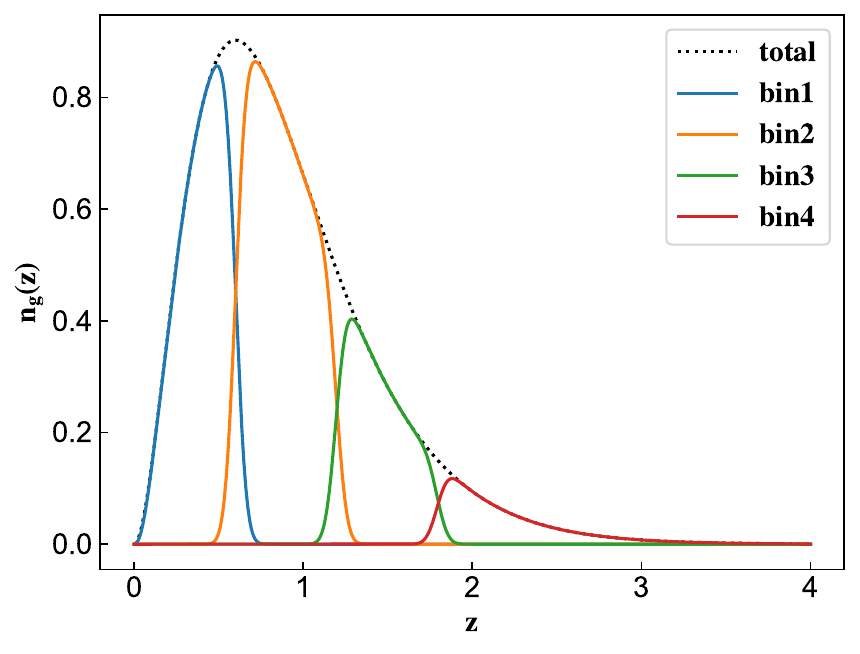}
    \includegraphics[width=0.49\linewidth]{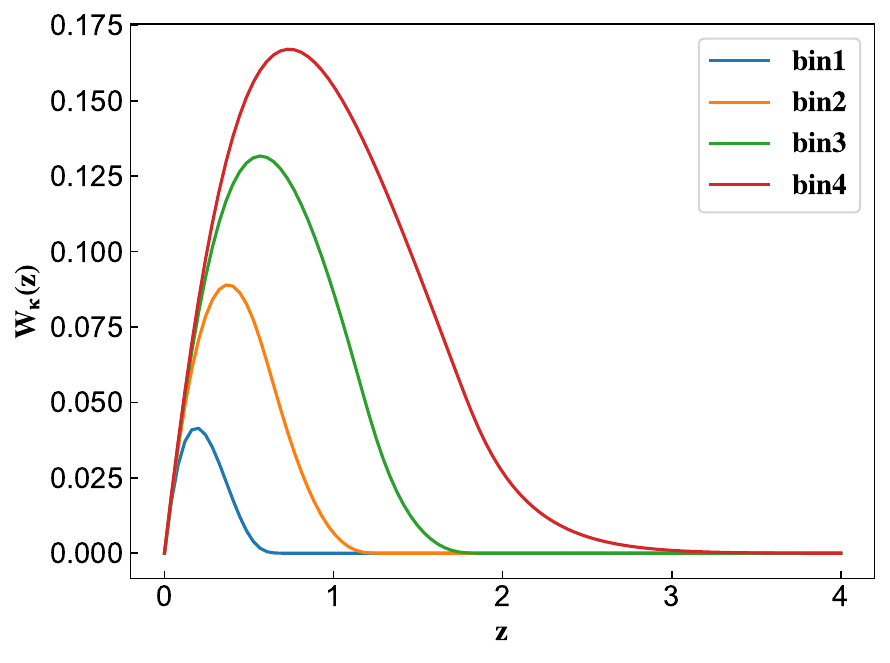}
    \caption{{\it Left panel:} The adopted galaxy redshift distributions in the CSST photometric survey. The black dotted curve shows the normalized total galaxy redshift distribution $n_{\rm g}(z)$, and the blue, orange, green, and red curves denote the galaxy redshift distribution in the four tomographic bins. {\it Right panel:} The weighting functions of the CSST weak lensing survey for the four tomographic bins. \label{fig: galaxy and shear kernal}}
\end{figure*}

\begin{figure*}
    \centering
    \includegraphics[width=0.95\linewidth]{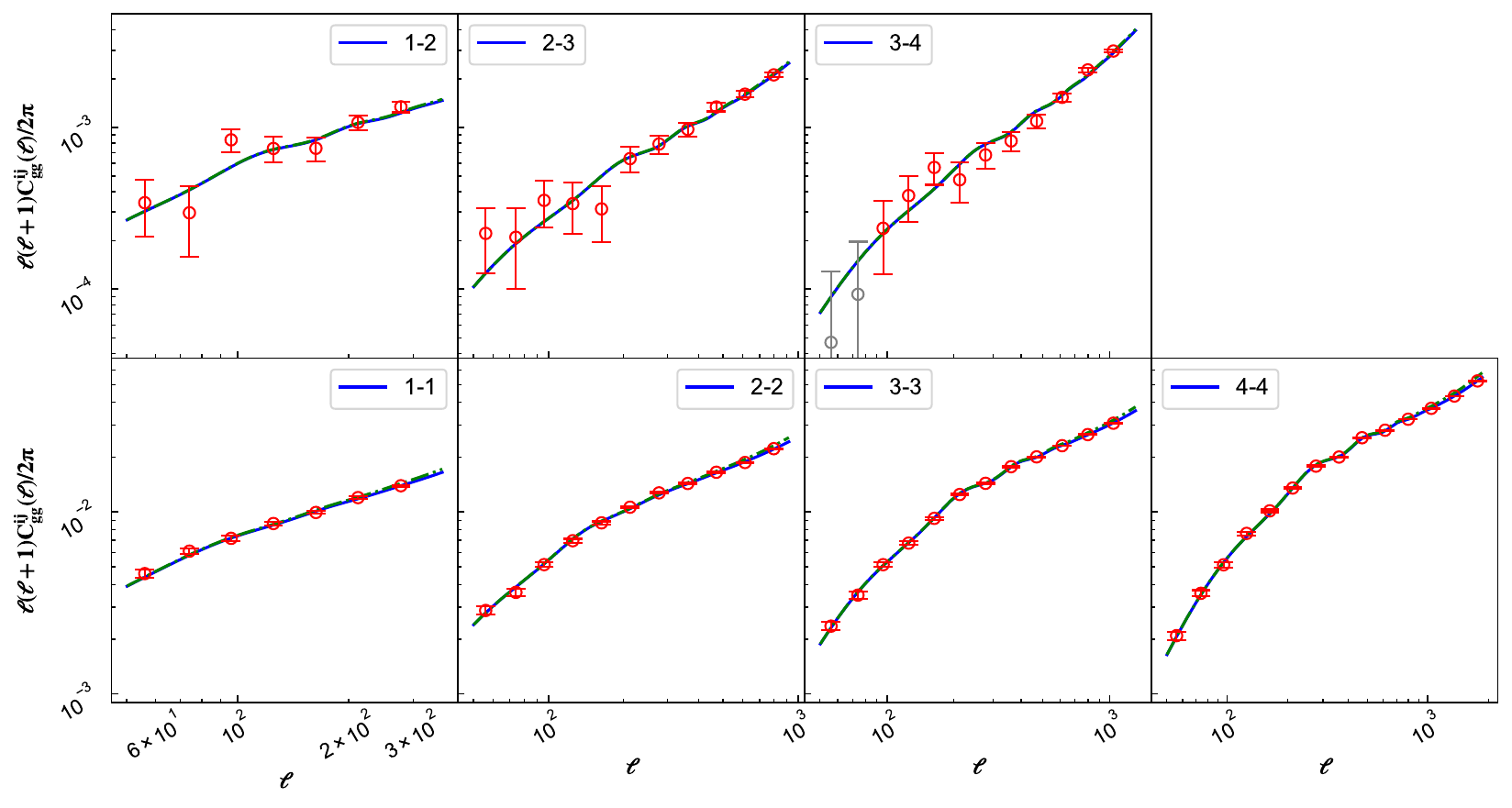}
    \caption{The theoretical curves and mock data of the galaxy angular power spectra in different tomographic bins for the CSST 3\(\times\)2pt analysis. The blue solid curves represent the fiducial theoretical predictions, and the green dashed curves denote the theoretical spectra for $m_{\rm PBH}=10^6$ $\rm M_{\odot}$ with \(f_{\mathrm{PBH}}=1\) as a reference. The red data points show the mock data used, while the gray data points indicate the excluded data with the signal-to-noise ratio (SNR) $<1$. Due to the flat sky assumption, Limber approximation and to avoid nonlinear effects, we have set lower and upper limits on \(l\) for different redshift tomography bins. The galaxy cross-power spectra of different tomographic bins with low amplitudes and small overlapping redshift ranges of $n_{\rm g}^i(z)$ are also removed. \label{fig:32pt Cgg}}
\end{figure*}

To estimate $n_{\rm g}^{i}(z)$, we assume that the total galaxy redshift distribution in the CSST photometric survey follows the function as \citep{Lin:2022aro}
\begin{equation}
    n_{\rm g} \left( z \right) \propto z^{2}e^{-z/z^{*}}.\label{equ: n of z}
\end{equation}
Here \(z^{*}=z_{\rm peak}/2=0.3\), where the distribution peak is \(z_{\rm peak}=0.6\). Following \cite{gong_cosmology_2019}, to utilize more information from the data, the redshift range is divided into four photometric tomographic bins, and we set the redshift bias $\Delta z^i=0$ and redshift scatter $\sigma^i_z=0.05$ as the fiducial values for each tomographic bin. These parameters are treated as free parameters during the fitting process. 

In the left panel of Figure~\ref{fig: galaxy and shear kernal}, we show the adopted galaxy redshift distributions. The black dotted curve denotes the total galaxy redshift distribution $n_{\rm g}(z)$, and it is normalized with $\int n_{\rm g}(z) dz=1$. The colored curves are the galaxy redshift distributions for the four tomographic bins. We find that the mean galaxy surface number densities in the four bins are $\bar{n}_{\rm g}^{i}= 7.9, 11.5, 4.6, 3.7\  {\rm arcmin}^{-2}$, respectively, and the corresponding total number density is $\bar{n}_{\rm g}=27.7\ {\rm arcmin}^{-2}$ which is consistent with the previous studies \citep[e.g.][]{gong_cosmology_2019}. In Figure~\ref{fig:32pt Cgg}, the blue solid curves denote the theoretical galaxy angular auto- and cross-power spectra for different tomographic bins in the CSST photometric survey without PBHs, and the green dashed curves denote the theoretical spectra for $m_{\rm PBH}=10^6$ $\rm M_{\odot}$ with \(f_{\mathrm{PBH}}=1\) as a reference.

\subsection{Shear power spectrum}

The shear auto- or cross-power spectra measured by the CSST for the $i$-th and $j$-th bins can be estimated by \citep[e.g.][]{huterer_systematic_2006}
\begin{equation}
     \widetilde{C}_{\gamma\gamma}^{i j}(\ell) =\left(1+ m_{i}\right) \left(1 + m_{j}\right) C_{\gamma\gamma}^{ij}(\ell)+ \delta_{i j} \frac{\sigma_{\gamma}^{2}}{\bar{n}_{\rm g}^{i}} + N_{\text{add},\gamma}^{ij},\label{equ: shear power spectra observed}   
\end{equation}
where \(\delta_{i j}\sigma_{\gamma}^{2}/\bar{n}_{\rm g}^{i}\) represents the shape shot noise term, \(\sigma_{\gamma}^2=0.04\) is the the shear variance \citep{gong_cosmology_2019}. The quantity \(m_i\) denotes the parameter for the multiplicative error in the $i$-th bin. The term \(N_{\text{add},\gamma}^{ij}\) is the additive error, which should be redshift and scale-dependent. For simplicity, we adopt a constant value of \(N_{\text{add},\gamma}^{ij}=N_{\text{add}}^{\gamma}\) for all redshift bins, and we set \(m_i=0\) and \(N_{\text{add}}^{\gamma}=10^{-9}\) as the fiducial values \citep{gong_cosmology_2019}. The shear signal power spectrum is \(C_{\gamma\gamma}^{ij}(\ell)\), which is given by \citep{Hildebrandt:2016iqg}
\begin{equation}
    C_{\gamma\gamma}^{i j} = P_{\kappa}^{i j}(\ell) + C_{\rm II}^{i j}(\ell) + C_{\mathrm{\rm GI}}^{i j}(\ell),\label{equ: shear signal power spectrum}
\end{equation}
where \(P_{\kappa}^{i j}(\ell)\) is the convergence power spectrum, which is the desired signal for cosmological analysis, \(C_{\rm II}^{i j}(\ell)\) is Intrinsic-Intrinsic power spectrum, which comes from the correlation between the intrinsic ellipticities of two galaxies. \(C_{\mathrm{GI}}^{i j}(\ell)\) is the Gravitational-Intrinsic power spectrum, which arises from the correlations between the gravitational shear of one galaxy and the intrinsic shape of another galaxy \citep{joachimi_galaxy_2015}.

\begin{figure*}
    \centering
    \includegraphics[width=0.95\linewidth]{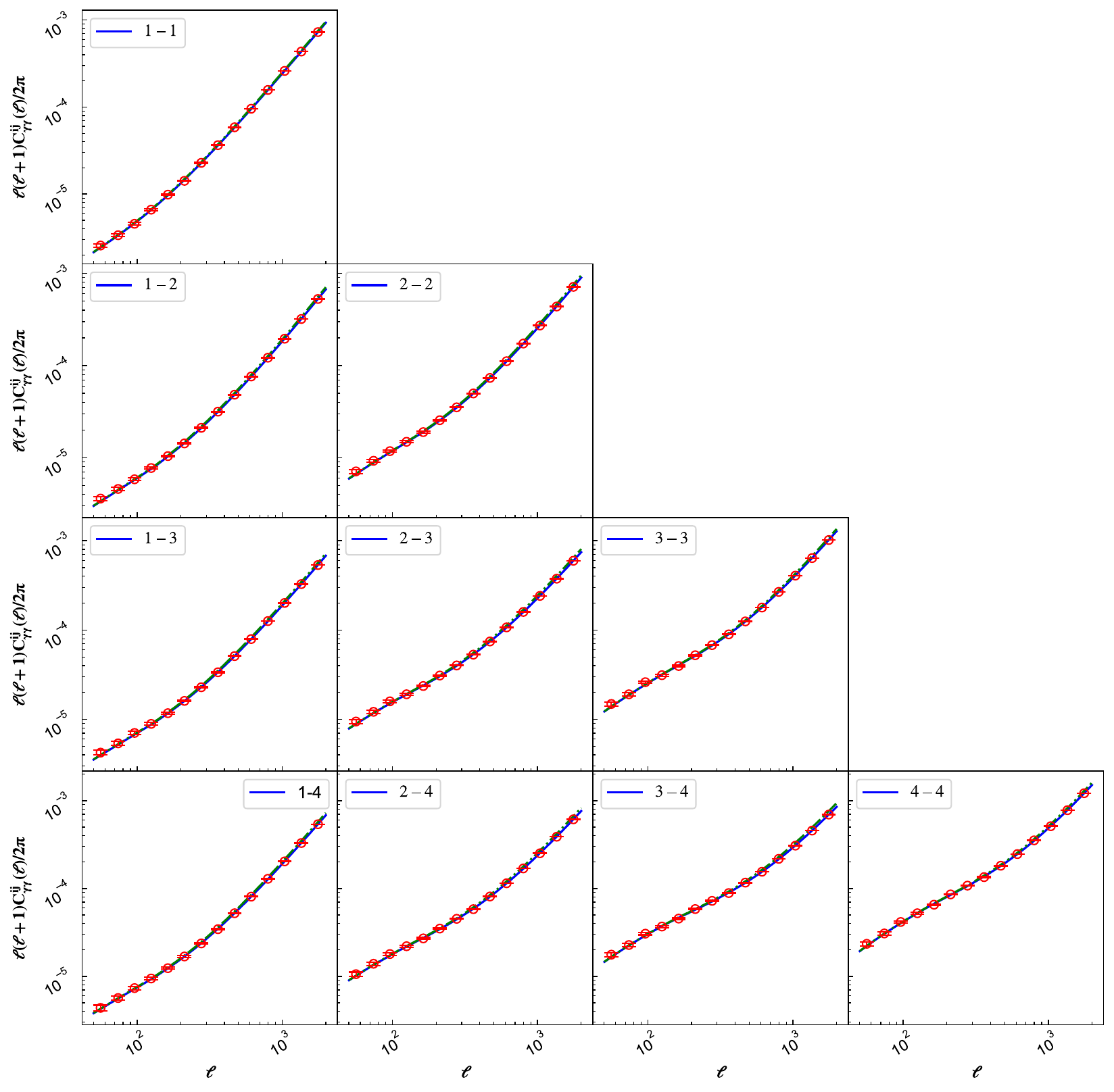}
    \caption{The  theoretical curves and mock data of the shear angular power spectra for the CSST 3\(\times\)2pt analysis. The blue solid curves are from the fiducial theoretical models, the green dashed curves denote the theoretical spectra for $m_{\rm PBH}=10^6$ $\rm M_{\odot}$ with \(f_{\mathrm{PBH}}=1\) as a reference. The red data points show the mock data used. Due to the flat sky assumption and Limber approximation, we have set lower limit on \(l\) for different redshift tomography bins.
    \label{fig:32pt Crr}}
\end{figure*}

\begin{figure*}
    \centering
    \includegraphics[width=0.93\linewidth]{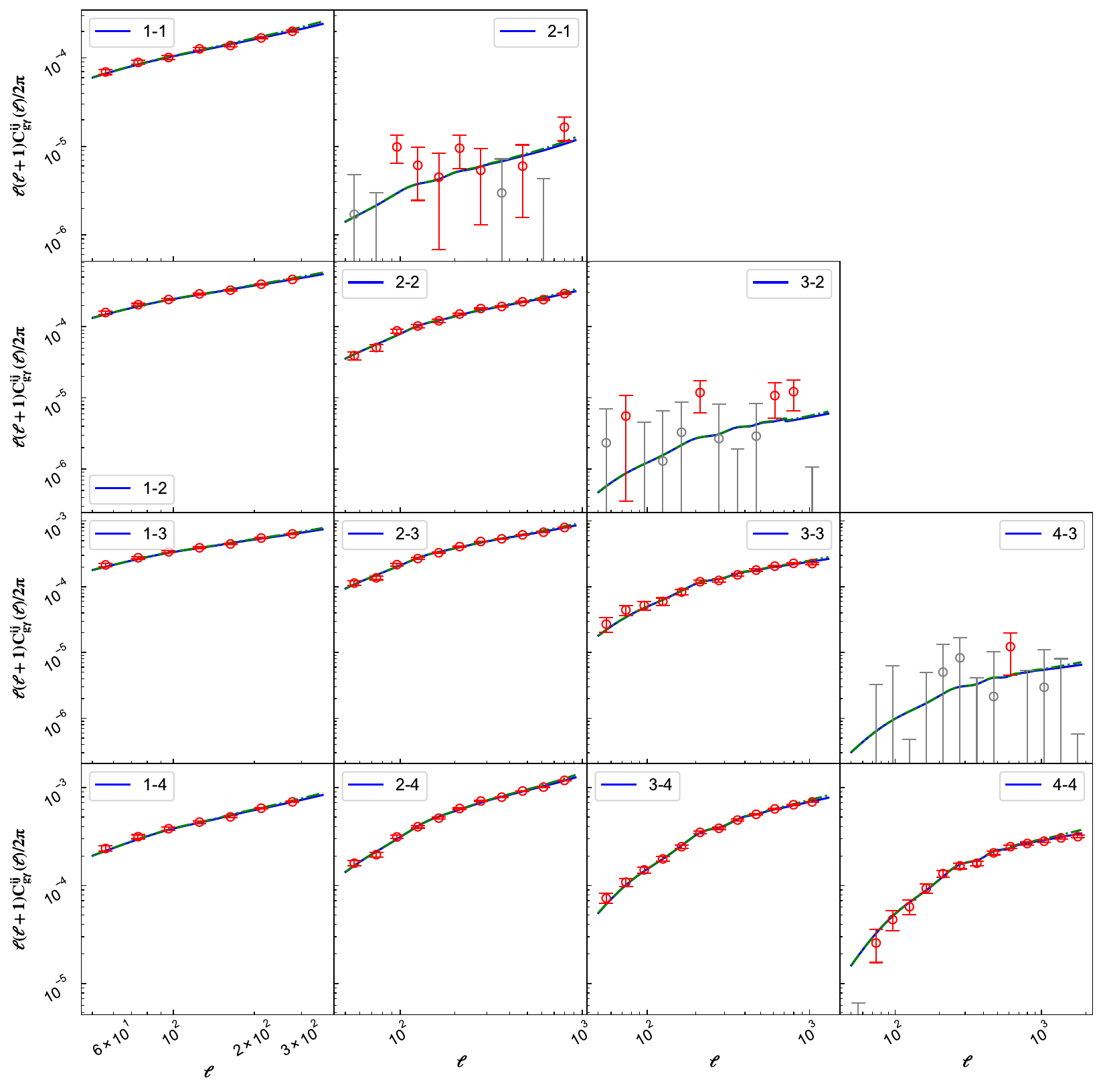}
    \caption{The theoretical curves and mock data of the galaxy-galaxy lensing angular power spectra in different tomographic bins for the CSST 3\(\times\)2pt analysis. The blue solid curves represent the fiducial theoretical predictions, the green dashed curves denote the theoretical spectra for $m_{\rm PBH}=10^6$ $\rm M_{\odot}$ with \(f_{\mathrm{PBH}}=1\) as a reference. The red data points show the mock data used, while the gray data points indicate the excluded data with the SNR $<1$. Due to the flat sky assumption, Limber approximation and to avoid nonlinear effects, we have set lower and upper limits on \(l\) for different redshift tomography bins. The galaxy-galaxy lensing cross-power spectra of different tomographic bins with low amplitudes and small overlapping redshift ranges of $n_{\rm g}^i(z)$ and $W_{\kappa}^i(z)$ are also removed. \label{fig:32pt Cgr}}
\end{figure*}
Assuming a flat sky and the Limber approximation \citep{Limber:1954zz}, the convergence power spectrum  \(P_{\kappa}^{i j}(\ell)\) is given by \citep{hu_joint_2004}
\begin{equation}
    P_{\kappa}^{ij}(\ell) = \int dz\, H(z)\frac{  W_{\kappa}^{i}(z) W_{\kappa}^{j}(z)}{c D_{A}^{2}(z)} P_{m}( \frac{\ell + \frac{1}{2}}{D_{\mathrm{A}}(z)}, z ),\label{equ: pkk}
\end{equation}
where \(W_{\kappa}^{i}(z)\) is the lensing kernel function of the $i$-th bin, and it can be estimated by
\begin{equation}
    W_{\kappa}^{i}(z) = \frac{3\Omega_{m} H_{0}^{2}D_{\mathrm{A}}(z)}{2cH(z)a}  \int_{z}^{\infty} dz^{\prime} \, n_{\rm g}^{i}(z^{\prime}) \frac{D_{\mathrm{A}}(z, z^{\prime})}{D_{\mathrm{A}}(z^{\prime})}.\label{equ: wkk}
\end{equation}
Here the integration limits are from \(z=0\) to 4. The right panel of Figure \ref{fig: galaxy and shear kernal} shows the lensing weighting kernels for the four bins. We can see that $W_{\kappa}^{i}(z)$ of the high-redshift bins can cover the low-redshift bins, which have wider distributions compared to $W_{\kappa}^{i}(z)$ of the low-redshift bins. 

The Intrinsic-Intrinsic power spectrum \(C_{\rm II}^{i j}(\ell)\) is given by \citep{Hildebrandt:2016iqg}
\begin{equation}
    \begin{aligned}
    C_{\rm II}^{ij} &=  \int dz \, H(z) \frac{F^{i}(z) F^{j}(z)}{cD_{\mathrm{A}}^{2}(z)}  P_{m} ( \frac{\ell + 1/2}{D_{\mathrm{A}}(z)}, z ),\\
    F^{i}(z) &= A_{\mathrm{IA}} C_{1} \rho_{\mathrm{c}} \frac{\Omega_{m}}{D(z)} n_{\rm g}^{i}(z)( \frac{1+z}{1+z_{0}} )^{\eta_{\mathrm{IA}}} ( \frac{L_{i}}{L_{0}} )^{\beta_{\mathrm{IA}}}.\label{equ: cii}
    \end{aligned}
\end{equation}
Here \(F^{i}(z)\) is the weighting kernel of the intrinsic alignment effect, the constant \(C_1=5\times10^{-14}h^{-2}M_{\odot}^{-1}{\rm Mpc}^3\), \(\rho_c\) represents the present critical density, \(D(z)\) is the linear growth factor, \(z_0\) and \(L_0\) are the pivot redshift and luminosity, and \(A_{\mathrm{IA}}\), \(\eta_{\mathrm{IA}}\) and \(\beta_{\mathrm{IA}}\) are free parameters. In this study, we adopt \(z_0=0.6\) and set the fiducial values  \(A_{\mathrm{IA}}=1\) and \(\eta_{\mathrm{IA}}=0\) \citep{Hildebrandt:2016iqg}. For simplicity, we do not consider the luminosity dependence and fix \(\beta_{\mathrm{IA}}=0\) \citep{gong_cosmology_2019,Lin:2022aro}.  

The Gravitational-Intrinsic power spectrum consists of two components, representing the cross-correlation between gravitational lensing and intrinsic alignment for the $i$-th and $j$-th bins, which can be estimated by
\begin{equation}  
        C_{\mathrm{GI}}^{ij} = \frac{1}{c} \int \mathrm{d} z \, H(z) D_{\mathrm{A}}^{-2}(z) \left[ W^{i}(z) F^{j}(z)+ W^{j}(z) F^{i}(z) \right] P_{m} \left( \frac{\ell + 1/2}{D_{\mathrm{A}}(z)}, z \right) .\label{equ: c gi}
\end{equation}
We find that the relationship \(C_{\gamma\gamma}>C_{\rm GI}>C_{\rm II}\) holds, and the effect of PBHs becomes significant when \(l\gtrsim10^3\). In Figure~\ref{fig:32pt Crr}, the blue solid curves denote the theoretical shear angular auto- and cross-power spectra for different tomographic bins in the CSST photometric survey without PBHs, and the green dashed curves denote the theoretical spectra for $m_{\rm PBH}=10^6$ $\rm M_{\odot}$ with \(f_{\mathrm{PBH}}=1\).
 
\subsection{Galaxy-galaxy lensing power spectrum}
The galaxy-galaxy lensing power spectrum represents the cross correlation between the galaxy and weak lensing. It is expressed as \citep{DES:2017myr}
\begin{equation}
     C_{\mathrm{g}\gamma}^{i j}(\ell) = C_{\mathrm{g}\kappa}^{i j}(\ell) + C_{\mathrm{gI}}^{i j}(\ell)+ N_{\text{add},g\gamma}^{ij} .\label{equ: c gr}
\end{equation}
The first term corresponds to the cross-correlation between galaxies and cosmic shear, and the second term is for the cross-correlation between galaxies and intrinsic alignment. The third term \(N_{\text{add},g\gamma}^{ij}\) is the additive error, which should be redshift and scale-dependent. For simplicity, we adopt a constant value of \(N_{\text{add},g\gamma}^{ij}=N_{\text{add}}^{g\gamma}\) for all redshift bins, and we set \(N_{\text{add}}^{g\gamma}=10^{-9}\) as the fiducial values \citep{gong_cosmology_2019}. The expressions for  \(C_{\mathrm{g}\kappa}^{i j}(\ell)\) and \(C_{\mathrm{gI}}^{i j}(\ell)\) are
\begin{equation}
    \begin{aligned}
     C_{\mathrm{g}\kappa}^{i j}(\ell) &= \int dz \, H(z) \frac{W_{\mathrm{g}}^{i}(z) W_{\kappa}^{j}(z)}{cD_{\mathrm{A}}^{2}(z)}   P_{m}( \frac{\ell + 1/2}{D_{\mathrm{A}}(z)}, z ),\\
     C_{\mathrm{g}\mathrm{I}}^{i j}(\ell) &= \int dz \, H(z) \frac{W_{\mathrm{g}}^{i}(z) F^{j}(z)}{cD_{\mathrm{A}}^{2}(z)} P_{m}( \frac{\ell + 1/2}{D_{\mathrm{A}}(z)}, z ).\label{eau:c_gk }
    \end{aligned}
\end{equation}

Since the cross-correlation between galaxy and cosmic shear \(C_{\mathrm{g}\kappa}^{i j}(\ell)\) exceeds that between galaxy and intrinsic alignment \(C_{\mathrm{gI}}^{i j}(\ell)\), \(C_{\mathrm{g}\kappa}^{i j}(\ell)\) dominates the total signal. We find that the effect of PBHs becomes significant when \(l\gtrsim10^3\). In Figure~\ref{fig:32pt Cgr}, the blue solid curves denote the theoretical galaxy-galaxy lensing power spectra for different tomographic bins in the CSST photometric survey without PBHs, and the green dashed curves denote the theoretical spectra for $m_{\rm PBH}=10^6$ $\rm M_{\odot}$ with \(f_{\mathrm{PBH}}=1\).

Note that the magnification bias is another important effect in the $3\times2$pt probe, and it may significantly affect the constraint results, especially for the Stage IV survey \citep{Thiele:2019fcu,Lorenz:2017iez,Euclid:2021rez}. However, since the CSST weak lensing measurement provides the main constraint power on the cosmological parameters as shown in Section~\ref{sec: result}, and the magnification bias mainly affect the galaxy auto and galaxy-galaxy lensing terms, we neglect this effect in our analysis for simplicity. We will study the impact of this effect in our future work.


\section{Mock Data}\label{sec: mock data}

\begin{table}
\centering
\footnotesize
\setlength{\tabcolsep}{2.5pt}
\setlength{\extrarowheight}{1pt}
\caption{The fiducial values, priors, best-fitting values and $1\sigma$ errors of the free parameters in the CSST photometric surveys. The uniform and Gaussian priors are denoted by \(U(a,b)\) and \(N ( \mu , \sigma)\), respectively, where $a$ and $b$ are the prior range and \(\mu\) and \(\sigma\) are the mean and standard deviation. The relative accuracy for each parameter is also provided in brackets. Here, \(fm=f_{\rm PBH}m_{\rm PBH}/M_{\odot}\).}
\label{tab:free parameters prior and result}
\begin{tabular}{ccccccc}
\toprule
\multirow{2}{*}{Parameter} & 
\multirow{2}{*}{\makecell{Fiducial \\ Value}} & 
\multirow{2}{*}{Prior} &
\multicolumn{3}{c}{Constraints by} \\
\cmidrule(lr){4-6}
& & & 
\makecell{Galaxy \\ Clustering} & 
\makecell{Weak \\ Lensing} & 
\makecell{$3\times2$pt} \\
\midrule
\multicolumn{6}{c}{\textbf{Cosmological Parameter}}\\
\(\Omega_m\) & 0.32 & U(0.2,0.4) & \(0.334^{+0.030}_{-0.031}\)(9.2\%) &\(0.314^{+0.024}_{-0.024}\)(7.6\%)  & \(0.313^{+0.011}_{-0.010}\)(3.3\%)\\
\(\Omega_b\) & 0.048 & U(0.01,0.09) & \(0.0505^{+0.0068}_{-0.0068}\)(13\%) &\(0.042^{+0.012}_{-0.016}\)(34\%) & \(0.0500^{+0.0028}_{-0.0028}\)(5.6\%)\\
\(h\) & 0.6727 & N(0.6727,0.0060) & \(0.6732^{+0.0057}_{-0.0058}\)(0.85\%) & \(0.6728^{+0.0060}_{-0.0061}\)(0.90\%) & \(0.6729^{+0.0055}_{-0.0058}\)(0.84\%)\\
\(n_s\) & 0.96 & U(0.85,1.15) & \(0.947^{+0.039}_{-0.038}\)(4.0\%) & \(0.930^{+0.048}_{-0.047}\)(5.1\%) & \(0.980^{+0.020}_{-0.022}\)(2.1\%)\\
\(w\) & -1 & U(-1.5,-0.8) & \(-1.10^{+0.20}_{-0.25}\)(21\%) & \(-1.22^{+0.23}_{-0.19}\)(17\%) & \(-1.20^{+0.15}_{-0.17}\)(13\%)\\
\(\sigma_8\) & 0.8 & U(0.7,0.9) & \(0.790^{+0.037}_{-0.031}\)(4.3\%) & \(0.801^{+0.020}_{-0.019}\)(2.5\%)& \(0.805^{+0.015}_{-0.012}\)(1.7\%)\\
\(\log_{10}(fm+1)\) & 0 & U[0,8) & \(<4.6\) & \(<4.1\) & \(<3.9\)\\
\cmidrule(r){1-6}
\multicolumn{6}{c}{\textbf{Baryonic Effect}} \\
\({\rm log}_{10}(T_{\rm AGN}/K)\) & 7.8 & U(7.0,8.3) & \(<8.0\) & \(7.52^{+0.22}_{-0.32}\)(3.6\%) & \(7.756^{+0.094}_{-0.093}\)(1.2\%)\\
\cmidrule(r){1-6}
\multicolumn{6}{c}{\textbf{Intrinsic Alignment}}\\  
\(A_{\rm IA}\) & 1 & U(-5,5) & --- & \(1.00^{+0.17}_{-0.13}\)(15\%) & \(0.968^{+0.041}_{-0.036}\)(4.0\%)\\
\(\eta_{\rm IA}\) & 0 & U(-5,5) & --- & \(0.11^{+0.67}_{-0.69}\) & \(0.144^{+0.085}_{-0.087}\)\\
\cmidrule(r){1-6}
\multicolumn{6}{c}{\textbf{Galaxy Bias}} \\
\(b^1\) & 1.252 & U(0,5) & \(1.221^{+0.071}_{-0.068}\) & --- & \(1.242^{+0.022}_{-0.024}\)\\
\(b^2\) & 1.756 & U(0,5) & \(1.777^{+0.069}_{-0.076}\) & --- & \(1.767^{+0.037}_{-0.036}\)\\
\(b^3\) & 2.260 & U(0,5) & \(2.31^{+0.10}_{-0.10}\) & --- & \(2.306^{+0.062}_{-0.061}\)\\
\(b^4\) & 3.436 & U(0,5) & \(3.49^{+0.18}_{-0.18}\) & --- & \(3.50^{+0.11}_{-0.11}\)\\
\cmidrule(r){1-6}
\multicolumn{6}{c}{\textbf{Photo-$z$ Calibration}} \\
\(\Delta z^1\) & 0 & U(-0.1,0.1) & \(-0.027^{+0.047}_{-0.040}\) & \(-0.004^{+0.023}_{-0.028}\)& \(-0.0074^{+0.0077}_{-0.0097}\)\\
\(\Delta z^2\) & 0 & U(-0.1,0.1) & \(-0.018^{+0.040}_{-0.035}\) & \(0.001^{+0.024}_{-0.023}\) & \(-0.0073^{+0.0081}_{-0.0010}\)\\
\(\Delta z^3\) & 0 & U(-0.1,0.1) & \(-0.011^{+0.041}_{-0.038}\) & \(-0.000^{+0.034}_{-0.031}\) & \(-0.008^{+0.011}_{-0.014}\)\\
\(\Delta z^4\) & 0 & U(-0.1,0.1) & \(-0.010^{+0.052}_{-0.046}\) & \(0.021^{+0.055}_{-0.053}\) & \(-0.016^{+0.017}_{-0.018}\)\\
\(\sigma^1_z/\sigma^1_{z,\rm fid}\) & 1 & U(0.5,1.5) & \(1.00^{+0.32}_{-0.33}\) & \(1.07^{+0.29}_{-0.34}\)& \(1.0^{+0.1}_{-0.1}\)\\
\(\sigma^2_z/\sigma^2_{z,\rm fid}\) & 1 & U(0.5,1.5) & \(1.02^{+0.32}_{-0.35}\) & \(1.09^{+0.30}_{-0.34}\) & \(0.957^{+0.060}_{-0.074}\)\\
\(\sigma^3_z/\sigma^3_{z,\rm fid}\) & 1 & U(0.5,1.5) & \(1.03^{+0.31}_{-0.34}\) & \(1.03^{+0.32}_{-0.34}\) & \(0.976^{+0.095}_{-0.093}\)\\
\(\sigma^4_z/\sigma^4_{z,\rm fid}\) & 1 & U(0.5,1.5) & \(0.87^{+0.52}_{-0.47}\) & \(0.99^{+0.32}_{-0.34}\) & \(0.82^{+0.25}_{-0.20}\)\\
\cmidrule(r){1-6}
\multicolumn{6}{c}{\textbf{Shear Calibration and Noise}} \\ 
\(m_1\) & 0 & U(-0.1,0.1) & --- & \(0.026^{+0.045}_{-0.052}\) & \(0.027^{+0.037}_{-0.036}\)\\
\(m_2\) & 0 & U(-0.1,0.1) & --- & \(0.018^{+0.044}_{-0.050}\) & \(0.019^{+0.032}_{-0.033}\)\\
\(m_3\) & 0 & U(-0.1,0.1) & --- & \(0.020^{+0.041}_{-0.047}\) & \(0.021^{+0.032}_{-0.033}\)\\
\(m_4\) & 0 & U(-0.1,0.1) & --- & \(0.017^{+0.041}_{-0.047}\) & \(0.023^{+0.032}_{-0.033}\)\\
\(N^g_{\rm sys}\) & \(10^{-8}\) & U\((0.5,1.5)\times10^{-8}\) & \(1.063^{+0.043}_{-0.042}\times10^{-8}\) & --- & \(9.93^{+0.26}_{-0.28}\times10^{-9}\)\\
\(N^\gamma_{\rm add}\) & \(10^{-9}\) & U\((0.5,1.5)\times10^{-9}\) & --- & \(9.987^{+0.027}_{-0.025}\times10^{-10}\) & \(1.0032^{+0.0021}_{-0.0024}\times10^{-9}\)\\
\(N^{g\gamma}_{\rm add}\) & 0 & U\((-1,1)\times10^{-8}\) & --- & --- & \(0.1^{+1.4}_{-1.4}\times10^{-11}\)\\
\bottomrule
\end{tabular}
\end{table}

To generate the mock data of the CSST $3\times2$pt probe, the covariance matrix is needed, which can be estimated by \citep{hu_joint_2004}
\begin{equation}
        \operatorname{Cov}[\tilde{C}_{x y}^{a b}(\ell) ,\tilde{C}_{mn}^{i j}(\ell^{\prime})]=\frac{\delta_{\ell \ell^{\prime}}}{(2\ell+1) f_{\text{sky}}\Delta\ell}[\tilde{C}_{xm}^{ai}(\ell)\tilde{C}_{yn}^{bj}(\ell)+\tilde{C}_{xn}^{aj}(\ell)\tilde{C}_{ym}^{bi}(\ell)] ,
\label{equ: covariance matrix}
\end{equation} 
where $a$, $b$, $i$, $j$ \(\in\) \{1, 2, 3, 4\} denote different tomographic bins, and $x$, $y$, $m$, $n$ \(\in\) \{\(g, \gamma\)\} denote galaxy or shear. $f_{\rm sky}\simeq0.42$ is the sky fraction for the CSST wide-field survey. The covariance matrix is decomposed using Cholesky decomposition. Then, it is multiplied by a standard normal random matrix. This yields a random displacement matrix with the same covariance. The random displacement data are added to the 3\(\times\)2pt theoretical predictions and forms the 3\(\times\)2 pt mock observed data. This procedure is trying to include the effect of the statistical uncertainties in real observations, and can help us to assess the impact on the parameter constraint. The mock data points of the CSST galaxy, weak lensing, and galaxy-lensing power spectra are shown in Figure~\ref{fig:32pt Cgg}, Figure~\ref{fig:32pt Crr}, and Figure~\ref{fig:32pt Cgr}, respectively. The error bars of the data points in these figures are derived from the diagonal components of the covariance matrix. As can be seen in these figures, since we have a large number of data with high SNR, it is highly impossible to have a systematic displacement or shift in the data points for the whole dataset to cause a large bias in the parameter constraint.

Since the flat sky assumption and Limber approximation are only valid at \(l\gtrsim50\) \citep{Limber:1954zz}, the data points of the angular power spectra at \(l<50\) are discarded in the fitting process. Besides, we only adopt the data points with ${\rm SNR}>1$ to improve the data quality and reduce computational time. The data points at small scales in the non-linear regime with $k>0.3\ {\rm Mpc}^{-1}h$ are also removed in the galaxy and galaxy-lensing power spectra, which corresponds to $l_{\rm max}={359, \,920,\, 1316,\, 1891}$ \citep{Lin:2022aro}. For the shear power spectra, we take $l_{\rm max}=2000$. In addition, we also discard the cross power spectra of different tomographic bins for galaxies and shear signals with low amplitudes in Figure~\ref{fig:32pt Cgg} and Figure~\ref{fig:32pt Cgr}, which have small overlapping redshift ranges for $n_{\rm g}^i(z)$, or $n_{\rm g}^i(z)$ and $W_{\kappa}^i(z)$, as shown in Figure~\ref{fig: galaxy and shear kernal}.

\section{Constraint and Results}\label{sec: result}
\subsection{Fitting method}

\begin{figure*}
    \centering
    \includegraphics[width=0.95\linewidth]{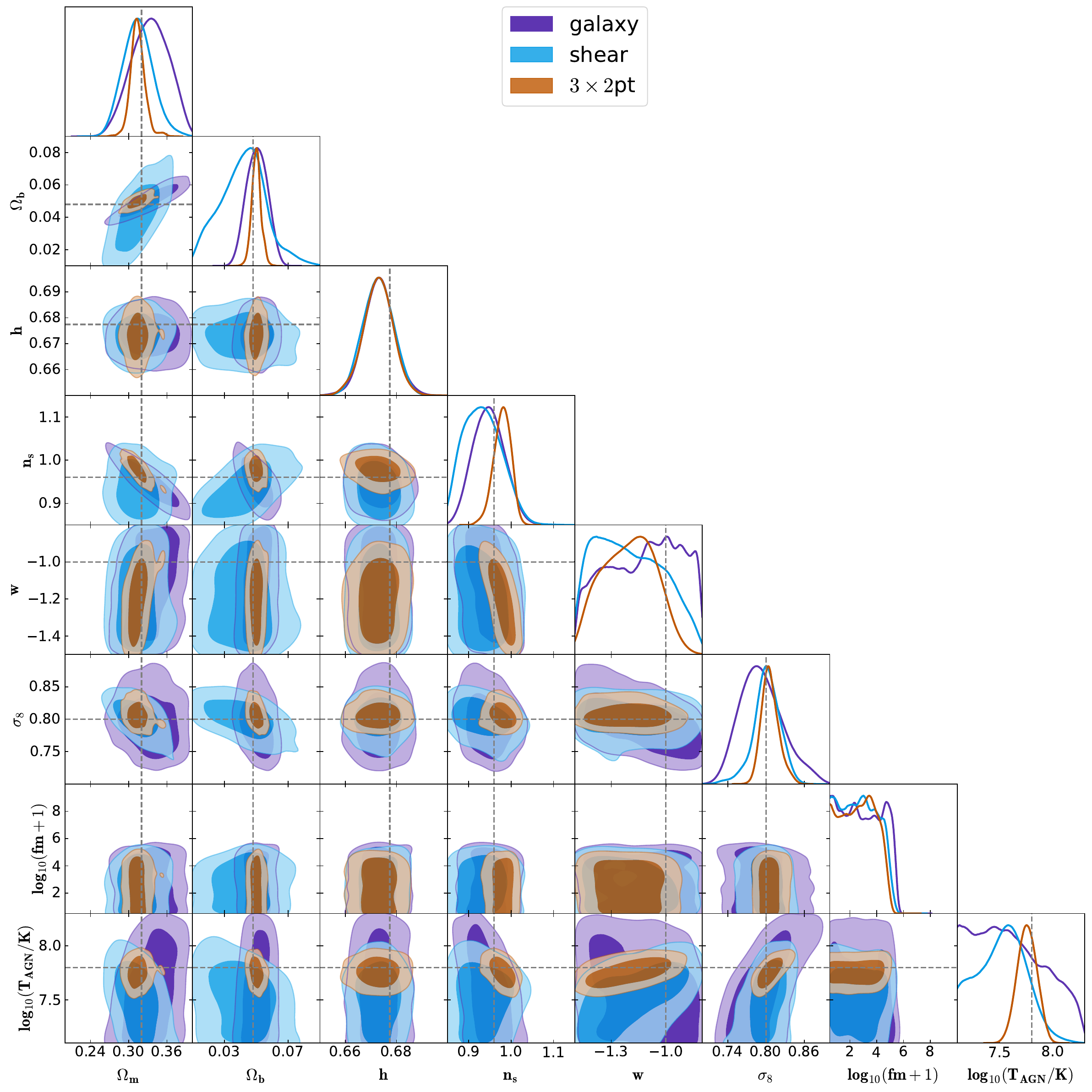}
    \caption{The contour maps with 1\(\sigma\) (68\%) and 2\(\sigma\) (95\%) CLs and 1D PDFs of the cosmological, PBH, and baryonic effect parameters for the CSST galaxy clustering (purple), cosmic shear (blue) and 3\(\times\)2pt (brown) surveys. The gray vertical and horizontal dashed lines indicate the fiducial values of these parameters. 
    \label{fig:cosmos parameter results}}
\end{figure*}

    We adopte the \(\chi^{2}\) method to fit the mock data, which is given by
\begin{equation}
    \chi^{2} = [\boldsymbol{D} - \boldsymbol{T}]^{T} \mathbf{C}\mathbf{o}\mathbf{v}^{-1} [\boldsymbol{D} - \boldsymbol{T}],
    \label{equ: chi^2}
\end{equation}
where \(\boldsymbol{D}\) represents the 3\(\times\)2pt data vector, \(\boldsymbol{T}\) is the corresponding theoretical prediction vector, and \(\mathbf{C}\mathbf{o}\mathbf{v}\) is the corresponding covariance matrix. The likelihood function follows  \(\mathcal{L} \sim \exp(-\chi^{2}/2)\).

The fitting process uses the  \href{https://emcee.readthedocs.io/en/stable/}{\tt emcee} \citep{foreman-mackey_emcee_2013} package, which implements an Markov Chain Monte Carlo (MCMC) ensemble sampler. In Table \ref{tab:free parameters prior and result}, we summarize the free parameters, their fiducial values and priors for the CSST galaxy clustering, weak lensing, and 3$\times$2pt surveys. 
Since the PBH parameter $f_{\rm PBH}m_{\rm PBH}$ spans nearly ten orders of magnitude, its logarithmic form is taken, i.e. ${\rm log}_{10}(fm+1)$ is used as a free parameter, where \(fm=f_{\rm PBH}m_{\rm PBH}/M_{\odot}\). The fiducial value is set to be \(fm=0\), which represents the case without PBHs. Note that the nuisance parameters including the systematic effects from baryonic feedback, intrinsic alignment, galaxy bias, photo-$z$ calibration, shear calibration, and additive noise, which cannot be fully eliminated in the data processing of shear and galaxy clustering measurements, are also jointly fitted with the cosmological parameters in our analysis to obtain more reasonable and reliable results. The priors of most of the free parameters are set to be flat, while we apply a Gaussian prior to the reduced/dimensionless Hubble parameter $h$ based on the $Planck$ result \citep{Planck:2018vyg} to effectively improve the constraint power on the other cosmological parameters, e.g. $\Omega_b$ and $\Omega_m$.

The MCMC implementation uses 150 walkers with 22,000 steps each to ensure the convergence. After burn-in and thinning process, approximately 10,000 chain points remain to illustrate the 1D probability distribution functions (PDFs) and contour maps of the free parameters.

\subsection{Constraint result}

In Figure~\ref{fig:cosmos parameter results}, the contour maps and 1D PDFs for the PBH, cosmological, and baryonic effect parameters have been shown. The 1$\sigma$ constraint result and the relative constraint accuracy for each parameter from the CSST galaxy clustering, weak lensing, and 3$\times$2pt surveys are listed in Table~\ref{tab:free parameters prior and result}.

We can find that, due to the large sky area and redshift coverage of the CSST photometric survey, the constraint accuracies of the PBH parameter product $f_{\rm PBH}m_{\rm PBH}$ can reach $<10^{3.9} M_{\odot}$ and $<10^{4.7} M_{\odot}$ for \(1\sigma\) and $2\sigma$ confidence levels (CLs), respectively. We also notice that the constraint on the product $f_{\rm PBH}m_{\rm PBH}$ from the CSST 3$\times$2pt analysis is similar to that from the CSST weak lensing survey, which indicates that the constraint power is mainly provided by the CSST shear measurement covering the small scales.

\begin{figure*}
    \centering
    \includegraphics[width=0.95\linewidth]{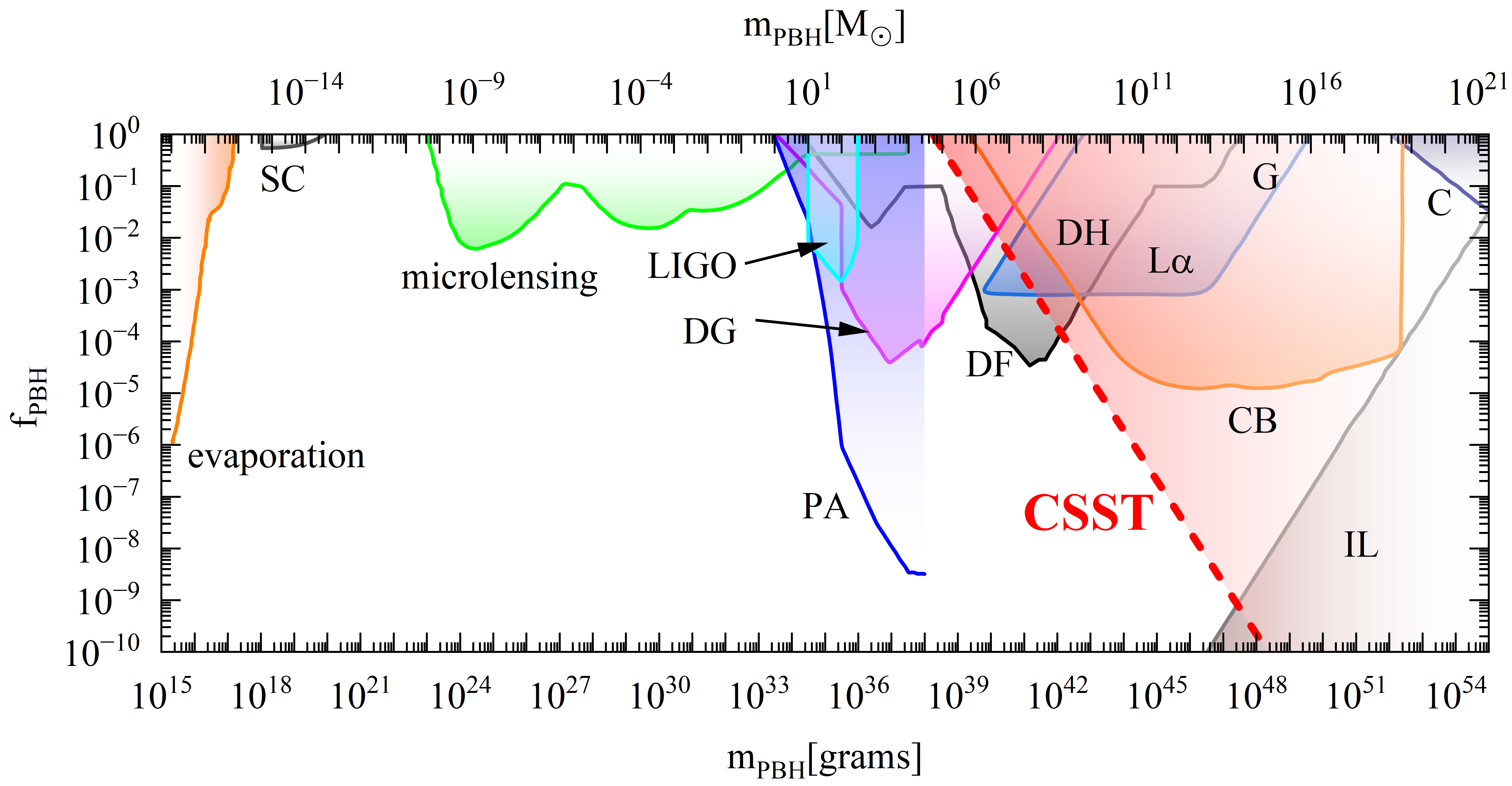}
    \caption{Comparison of the constraints on the  monochromatic mass function \(f_{\rm PBH}(m_{\rm PBH})\). The filled regions represent the excluded parameter space. The red dash line is derived from the \(2\sigma\) upper limit (\(f_{\rm PBH}m_{\rm PBH}<10^{4.7}M_{\odot}\)) given by this work using the CSST \(3\times2\)pt analysis. The solid lines indicate various previous constraints, including the limits from evaporation (orange), microlensing (green), Planck measurements of CMB distortions (PA, blue), halo dynamical friction (DF, black), heating of stars in the Galactic disk (DH, black), galaxy tidal distortions (G, black), CMB dipole (CMB, navy blue), and incredulity limits (IL, burgundy) \citep{carr_primordial_2020}. Additional constraints come from PBH capture by stars (SC, gray)\citep{Esser:2025pnt}, LIGO observation of the merger rate of PBH binaries (LIGO, cyan) \citep{ali-haimoud_merger_2017}, PBH in the halos of dwarf galaxies (DG, magenta) \citep{Pilipenko:2025qsf}, Lyman-\(\alpha\) forest (L\(\alpha\), sky blue) \citep{Ivanov:2025pbu}, and the combination of CMB and BAO (CB, orange) \citep{Gerlach:2025vco}. 
    \label{fig:all result}}
\end{figure*}

In Figure \ref{fig:all result}, we show the comparison of the constraint on $f_{\rm PBH}$ as a function of $m_{\rm PBH}$ with other results, based on our $2\sigma$ upper limit with $f_{\rm PBH}m_{\rm PBH}<10^{4.7}M_{\odot}$. The filled regions represent the excluded parameter space. In the mass range $m_{\rm PBH}\gtrsim10^8M_{\odot}$, we can see that the CSST \(3\times2\)pt analysis can provide tighter constraints than other current methods,  such as L\(\alpha\), CB, DH, and CMB, especially in the mass range \(10^{14} M_{\odot}<m_{\rm PBH}<10^{18}M_{\odot}\) where currently there is only one effective constraint. Therefore, it demonstrates strong potential for the future studies of massive PBHs using the CSST $3\times2$pt analysis.

For the cosmological parameters,  e.g. $\Omega_m$, $\sigma_8$, and $w$, we find that the CSST 3$\times$2pt analysis can provide the constraint accuracies $\sim3.3\%$, $1.7\%$, and $13\%$, respectively, which are much more stringent than those given by the current similar surveys, e.g. DES \citep{DES:2021wwk}. Compared with the previous CSST \(3\times2\)pt forecast \citep[e.g.][]{Lin:2022aro,Lin:2023yso}, we notice that the values of relative constraint accuracies of these parameters are larger. This is mainly because that we have included more parameters in this analysis, such as the noise parameters $N^g_{\rm sys}$, $N^\gamma_{\rm add}$, and $N^{g\gamma}_{\rm add}$.

In addition, we find that the systematical parameters are also correctly constrained in the CSST photometric surveys, and their fiducial values are within the $1\sigma$ CL, such as the parameters of baryonic effect ${\rm log}_{10}(T_{\rm AGN}/K)$, intrinsic alignment $A_{\rm IA}$ and $\eta_{\rm IA}$, galaxy bias in each redshift bin $b_{\rm g}^i$, photo-$z$ calibration in each redshift bin $\Delta z^i$ and $\sigma_z^i$, shear calibration in each redshift bin $m_i$, and noise terms $N^g_{\rm sys}$, $N^\gamma_{\rm add}$, and $N^{g\gamma}_{\rm add}$. A detailed discussion of these parameters is provided in appendix \ref{appendix}.

\section{Summary}\label{sec: summary}
In this work, we forecast the constraint on the product of the PBH CDM fraction and mass, i.e. $f_{\rm PBH}m_{\rm PBH}$, and other cosmological parameters using the CSST 3$\times$2pt analysis.  The modeling incorporates PBH isocurvature perturbations and ordinary $\Lambda$CDM adiabatic perturbations to construct the total matter power spectrum and then generate the mock data of the angular power spectra of galaxy clustering, weak lensing, and galaxy-galaxy lensing based on the design of the CSST photometric survey. The systematic parameters of baryonic effect, intrinsic alignment, galaxy bias, photo-$z$ calibration, shear calibration, and noise terms, are also considered and jointly constrained in the MCMC fitting process.

We find that the CSST \(3\times2\)pt analysis can provide tight constraints on the PBH parameter with $f_{\rm PBH}m_{\rm PBH}<10^{3.9}M_{\odot}$ and $<10^{4.7}M_{\odot}$ for \(1\sigma\) and $2\sigma$ CLs, respectively. It excludes significant portions of the parameter space in the mass range $m_{\rm PBH}>10^8 M_{\odot}$, especially for \(10^{14} M_{\odot}<m_{\rm PBH}<10^{18} M_{\odot}\) where the existing methods lack of sensitivity. We also notice that this constraint is mainly contributed by the CSST weak lensing measurement, which can provide more information at small scales. Besides, the constraints on the cosmological parameters, such as $\Omega_m$, $\sigma_8$, and $w$ can achieve much higher precision compared to the current photometric surveys, and the systematic parameters are also well-constrained. Our results validates the potential of the upcoming or ongoing Stage \Romannum{4} surveys for exploring the LSS evolution of the Universe, and indicates that the CSST 3$\times$2pt analysis can be a powerful tool for investigating the properties of PBHs.

\acknowledgments
D.A.H and Y.G. acknowledge the support from the CAS Project for Young Scientists in Basic Research (No. YSBR-092), and National Key R\&D Program of China grant Nos. 2022YFF0503404 and 2020SKA0110402. X.L.C. acknowledges the support of the National Natural Science Foundation of China through grant Nos. 11473044 and 11973047 and the Chinese Academy of Science grants ZDKYYQ20200008, QYZDJ- SSW-SLH017, XDB 23040100, and XDA15020200. This work is also supported by science research grants from the China Manned Space Project with grant Nos. CMS-CSST-2025-A02, CMS-CSST-2021-B01, and CMS-CSST-2021-A01.

\appendix

\section{constraint results of systematic parameters}\label{appendix}
\begin{figure}
    \centering
    \includegraphics[width=0.6\linewidth]{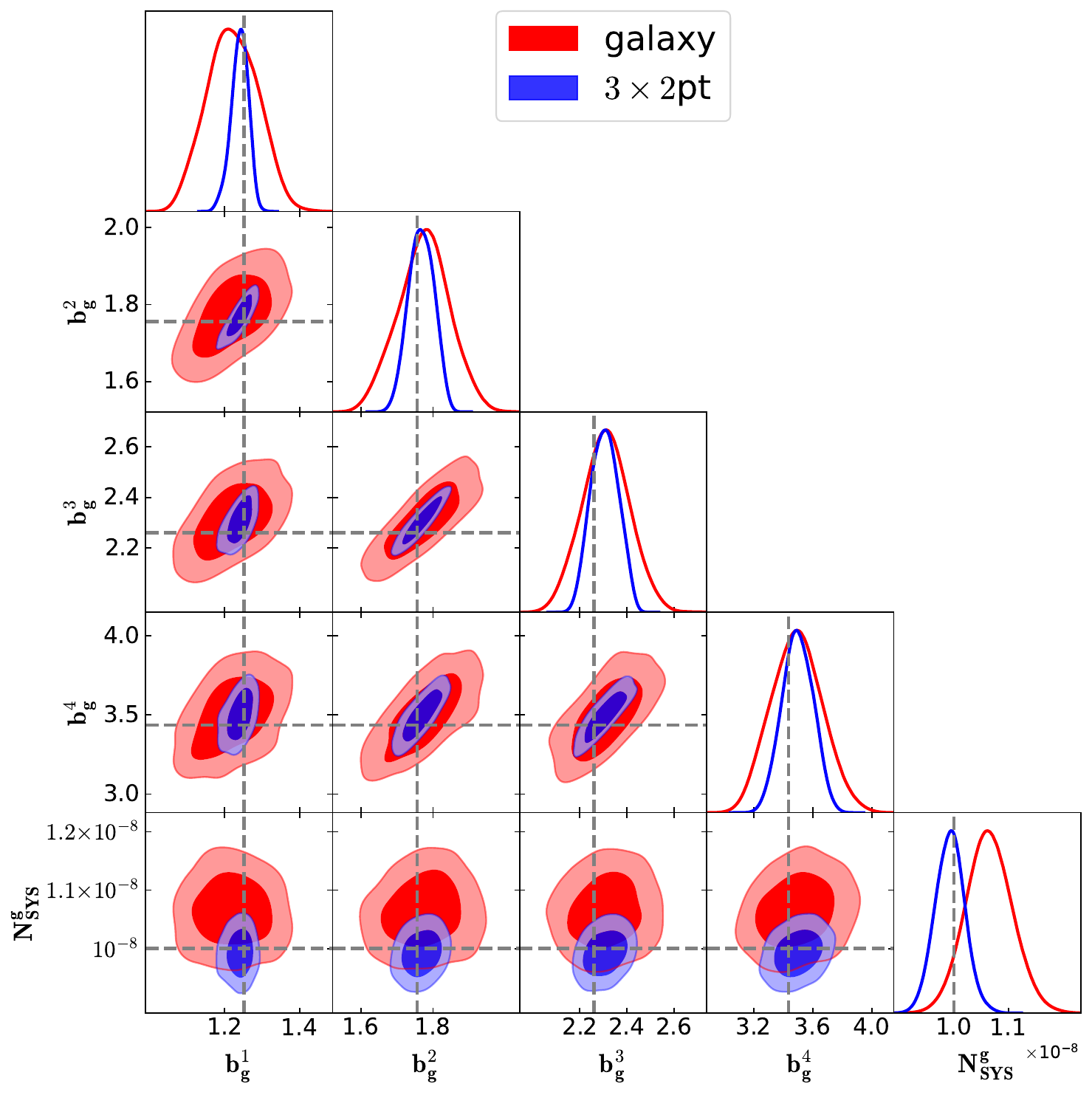}
    \caption{The contour maps with 1\(\sigma\) and 2\(\sigma\) CLs and 1D PDFs of galaxy bias and systematic noise for the CSST galaxy clustering (red) and 3\(\times\)2pt surveys (blue). 
    \label{fig: galaxy_parameter_results}}
\end{figure}
\begin{figure}
    \centering
    \includegraphics[width=0.6\linewidth]{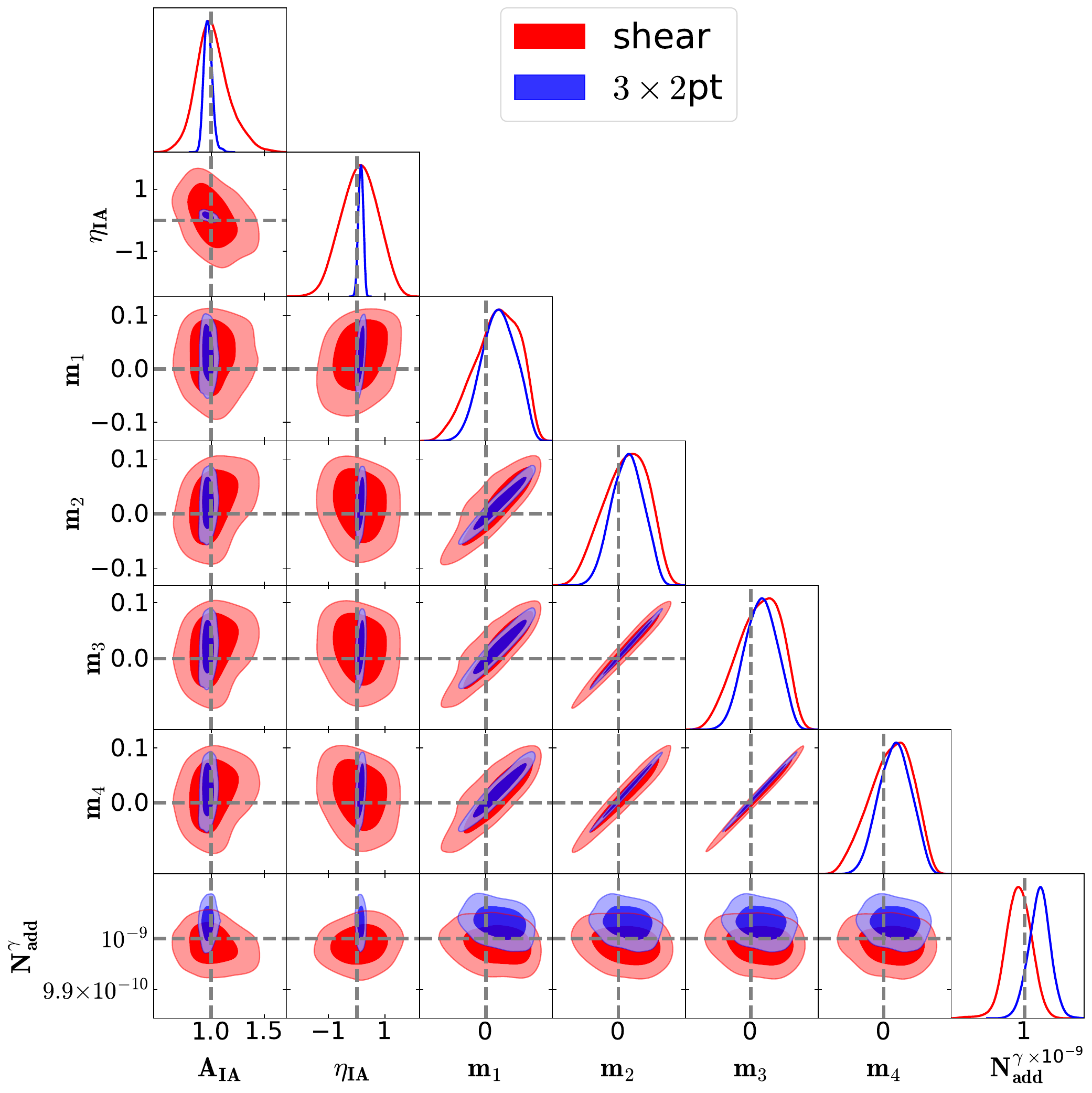}
    \caption{The contour maps with 1\(\sigma\) and 2\(\sigma\) CLs and 1D PDFs of intrinsic alignment, multiplicative and additive errors for the CSST shear (red) and 3\(\times\)2pt surveys (blue). The gray vertical and horizontal dashed lines indicate the fiducial values of these parameters.
    \label{fig: shear_parameter_results}}
\end{figure}

\begin{figure*}
    \centering
    \includegraphics[width=0.93\linewidth]{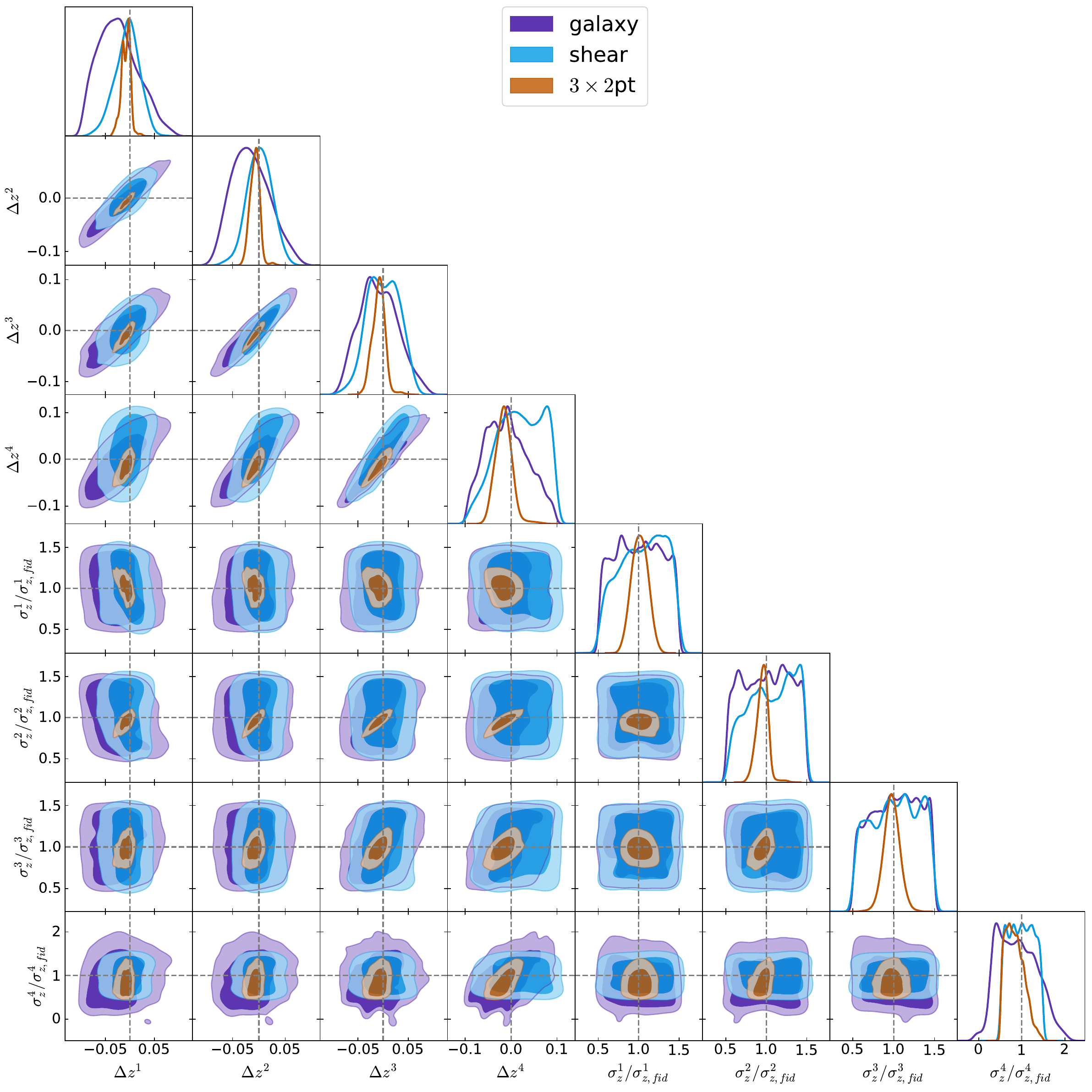}
    \caption{The contour maps with 1\(\sigma\) and 2\(\sigma\) CLs and 1d PDFs of the systematic parameters in the photo-$z$ calibration for the CSST galaxy clustering (purple), cosmic shear (blue) and 3\(\times\)2pt (brown) surveys. The gray vertical and horizontal dashed lines indicate the fiducial values of these parameters. 
    \label{fig: redshift sys parameter}}
\end{figure*}

The CSST photometric survey also can effectively constrain the systematic parameters in the model. By including the galaxy clustering, shear, and galaxy-galaxy lensing measurements in the $3\times2$pt probe, the degeneracies between the cosmological and systematic parameters can be effectively broken. The systematic parameters from the baryonic effect, intrinsic alignment, galaxy bias, and photo-$z$ calibration can be accurately constrained. Note that applying the matter power spectrum with baryonic feedback in the calculations of galaxy clustering and galaxy-galaxy lensing probes may lead to an overly stringent constraint on the parameter ${\rm log}_{10}(T_{\rm AGN}/K)$ in the $3\times2$pt analysis. However, since we have removed the data points at small scales with $k>0.3\ {\rm Mpc}^{-1}h$ in the galaxy clustering and galaxy-galaxy lensing power spectra, where the baryonic feedback becomes significant, 
 the constraint power on ${\rm log}_{10}(T_{\rm AGN}/K)$ is effectively limited for these two probes. As shown in Figure \ref{fig:cosmos parameter results}, we can see that the constraint on parameter ${\rm log}_{10}(T_{\rm AGN}/K)$ from the galaxy-only case is relatively weak, and it can only provide an upper limit of ${\rm log}_{10}(T_{\rm AGN}/K)<8.0$, whereas the \(3\times2\)pt can constrain ${\rm log}_{10}(T_{\rm AGN}/K)$ more effectively with an accuracy of $\sim$1\%. In Figure \ref{fig: galaxy_parameter_results} and \ref{fig: shear_parameter_results}, we show the contour maps and 1D PDFs of the parameters of galaxy bias, intrinsic alignment, multiplicative error, and noise terms in the CSST galaxy clustering, shear, and 3$\times$2pt measurements. Compared to the galaxy clustering measurement, the \(3\times2\)pt analysis can improve the constraint accuracies of the galaxy bias and shot noise by a factor of \(\sim2\). For the intrinsic alignment parameters, the \(3\times2\)pt method demonstrates a significant advantage in constraining \(A_{\rm IA}\), improving the constraint accuracy by a factor of \(\sim5\). In Figure \ref{fig: redshift sys parameter}, we show the contour maps and 1D PDFs of the systematic parameters in photo-$z$ calibration, i.e. $\Delta z^i$ and $\sigma_z^i$. Compared to the results from the CSST galaxy clustering only and shear only measurements, the CSST \(3\times2\)pt analysis can improve the constraint accuracy by a factor of \(\sim3\).

\bibliographystyle{JHEP}
\bibliography{reference}

\end{document}